\documentclass[12pt]{article}
\usepackage{amsmath,amsbsy,amssymb,graphics}
\usepackage{graphicx,epsfig}
\def\beq{\begin{equation}}
\def\eeq{\end{equation}}
\def\be{\begin{equation}}
\def\ee{\end{equation}}
\def\bea{\begin{eqnarray}}
\def\eea{\end{eqnarray}}
\def\nnb{\nonumber}

\newcommand{\gsim}{\lower.7ex\hbox{$\;\stackrel{\textstyle>}{\sim}\;$}}
\newcommand{\lsim}{\lower.7ex\hbox{$\;\stackrel{\textstyle<}{\sim}\;$}}

\begin{document}
 
\begin{center}
 \vspace{0.2cm}
 {\Large \bf Strong CP problem and spontaneous generation of  CP violating phase in CKM matrix}

\vspace{0.6cm} {\large \bf Wei Liao}

\vspace{0.3cm} {
 Institute of Modern Physics, School of Science,\\
 East China University of Science and Technology, \\
 130 Meilong Road, Shanghai 200237, P. R. China
 }
\end{center}
 \vskip 0.2cm
 \begin{abstract}
 \vskip 0.2cm
 We show that in a Complementary two-Higgs doublet model(C2HDM) the CP violating phase in the CKM matrix can 
 be generated spontaneously, dangerous FCNC can be naturally suppressed and the strong CP problem can also be 
 avoided.  The two Higgs doublets in the model are complementary in the sense that none of them is enough to describe 
 masses of a given type of quarks. We find that the strength of FCNC is suppressed by the strength of  Yukawa couplings 
 of the first generation quark and the tree-level FCNC is sufficiently small. Using an explicit example, we show that radiative 
 correction to the assumed Yukawa couplings can modify the discussion about the strong $\theta$.  The correction to the
 strong $\theta$ is estimated to be less than around $10^{-12}\sim10^{-10}$ which can be tested in future experiment.
 
\end{abstract}



\section{\large Introduction}\label{sec1}

One of the deep mysteries of particle physics is the origin of CP violation.
On one hand, CP symmetry is found to be broken in flavor changing processes of
$K$ and $B$ mesons. CP violating phenomena so far measured are
successfully explained by the CP violating phase in the complex Cabibbo-Kobayashi-Maskawa(CKM) 
matrix in the Standard Model(SM) . On the other hand,  the strong CP phase $\theta$ in 
\bea
\Delta {\cal L}=\frac{\alpha_s}{8\pi} \theta G_{\mu \nu} {\widetilde G}^{\mu \nu} \label{strongtheta1},
\eea
another possible source of CP violation in SM, has not been observed in experiment. 
On the contrary, this strong $\theta$ is found to be  $\theta \lsim 10^{-10}\sim 10^{-9}$, in measurements 
of Electric Dipole Moment(EDM) of neutron, mercury etc.~\cite{RPP}

The problem is quite challenging in view of the fact that the CP violating phase in the CKM matrix arises from
complex Yukawa couplings of quarks. These complex Yukawa couplings are natural to have 
non-zero flavor diagonal phases which can contribute to the physical strong $\theta$.  
More specifically, after spontaneous breaking of the $SU(2)_L\times U(1)_Y$ gauge symmetry, the mass terms of quarks in the SM are generated as
 \bea
M^{u,d}=Y^{u,d} ~v,
\eea
where $Y^{u,d}$  is the Yukawa coupling of up(down)-type quarks, $v=246/\sqrt{2}$ GeV the vacuum expectation value of the Higgs doublet in the SM.
Performing re-definitions of  left-handed and right-handed fields separately and diagonalizing the mass terms,
one can get the CKM matrix in charged current interaction of left-handed quarks.
 Meanwhile, the chiral $U(1)$ part of the field-redefinition would transform the $\theta$ term so that 
 the presence of these complex mass terms  or Yukawa terms  would give a contribution to the physical strong $\theta$
 \bea
 \theta=\theta_0 +\textrm{arg}(\det(M^u M^d)), \label{strongtheta2}
 \eea
where $\theta_0$ is the $\theta$ term before receiving correction. One would naturally expect the second term in (\ref{strongtheta2}) is not zero
if $Y^{u,d}$(or $M^{u,d}$) are complex matrices. So a very large fine-tuning between the two terms in (\ref{strongtheta2}) is required
to achieve a value of $\theta$  as small as  $\lsim 10^{-9}$.  In particular, a large fine-tuning seems un-avoidable if CP symmetry is
broken explicitly as in the SM.~\footnote{If making an extra assumption that there is no flavor-diagonal phase in complex Yukawa couplings, 
one can ignore this fine-tuning problem. 
In this case, flavor non-diagonal phase in CKM matrix can still contribute to
strong $\theta$ through radiative correction. But the first nonzero correction appears in 4th order in loop and
is of order $10^{-16}$\cite{Ellis-Gaillard}. We do not study this case concerning the CP violation in the SM.} 
This is the so-called strong CP problem~\cite{strongCP}.

One approach to understand the origin of CP violation is spontaneous breaking of CP symmetry~\cite{TDLee1}.
In this approach, CP symmetry is exact and the $\theta$ term is zero before the symmetry is broken spontaneously.
So spontaneous breaking of CP symmetry is a possible solution to the strong CP problem~\cite{Early,Early2,Early3}, and the
strong $\theta$ can be calculable in some new physics models.
Moreover, it was shown by some authors that the CP violating phase in the CKM matrix can be generated 
spontaneously~\cite{TDLee2, Chen et al}.  Although this is a very interesting approach to understand the origin of CP violation,
it's not straightforward to see whether the complex quark mass matrices generated in this kind of model
can still give a zero contribution to the $\theta$ term after the CP violating phase in the CKM matrix  is generated spontaneously.

In this paper, we are going to pursue the idea of spontaneous generation of flavor non-diagonal CP violating phase in the CKM matrix
and study the strong $\theta$ term in this approach.
We will show that the CP violating phase in the CKM matrix and a zero or very small strong $\theta$ can be
obtained simultaneously in a model of spontaneous generation of CP violation.  
First, we will show, using an explicit toy model, that the flavor non-diagonal 
CP violating phase in the CKM matrix can be generated spontaneously.  Since more than one Higgs doublets are needed in order to implement 
spontaneous CP violation, Yukawa couplings can be very complicated in general and dangerous  Flavor Changing Neutral Current(FCNC) 
processes could be generated. We show that there are some cases for which
FCNC processes can be naturally suppressed. Since CP symmetry is exact before spontaneous breaking,
the initial strong $\theta_0$ is zero. We show that the strong $\theta$ can still be zero in such kind of model
even when the CP violating phase in the CKM matrix is generated spontaneously.
We also study radiative corrections to Yukawa couplings and check the robustness of the 
above statement against possible radiative corrections.
We find that $\theta$ is smaller than around $10^{-12}\sim 10^{-10}$ in a particular model.

\section{\large Spontaneous generation of flavor changing CP violating phase} \label{sec2}
 In this section we show that the CP violating phase in the CKM matrix can be generated spontaneously.
 We assume that CP symmetry is exact before spontaneous symmetry breaking. 
 So Yukawa couplings are real and initial strong $\theta_0$ is zero.
  
Spontaneous CP violation in general involves more than one Higgs doublets, 
 e. g.   the two Higgs doublets $\phi_1$ and $\phi_2$. A general Lagrangian with two Higgs doublets which can give rise to
 a spontaneous generation of CP violating phase has been discussed in literature, e.g. in a recent
 paper \cite{MZH}.  In the present article we are not going to elaborate on this Lagrangian. 
 Instead, we assume that a spontaneous breaking of CP symmetry and a suitable CP violating phase can be achieved 
 with a suitable Lagrangian.
 We assume that $\phi_1$ and $\phi_2$, 
 both having hypercharge $-{1 \over 2}$,
 develop vacuum expectation values after spontaneous symmetry breaking
 \bea
 \langle \phi_1 \rangle= (v_1,0)^T, ~\langle \phi_2 \rangle= (v_2,0)^T, \label{vacuum}
 \eea
 where $v_1$ and $v_2$ are complex in general and they satisfy $|v_1|^2+|v_2|^2=v^2$.
 $\phi_1$ and $\phi_2$ can both couple to quarks: 
 \bea
-\Delta{\cal L}={\bar Q} Y^u_1 \phi_1 u_R+{\bar Q}Y^u_2 \phi_2 u_R+{\bar Q}Y^d_1 {\tilde \phi}_1 d_R+{\bar Q} Y^d_2{\tilde \phi}_2 d_R,
\label{yukawa2} 
 \eea
 where $Q$ is the field of the left-handed quark doublet, $u_R$ and $d_R$ the fields of right-handed up-type and down-type quarks. 
 ${\tilde \phi}_{1,2}=i\sigma^2 \phi^*_{1,2}$.
 Flavor indices have been suppressed in (\ref{yukawa2}). 
 $Y^{u,d}$ in (\ref{yukawa2}) are all real matrices so that CP symmetry is not broken explicitly.
 After spontaneous symmetry breaking the mass matrices of quarks are obtained as 
 \bea
 M^u= Y^{u}_1 v_1 +Y^{u}_2 v_2, \label{mass2}
 \eea
 and 
 \bea
 M^d= Y^{d}_1 v^*_1 +Y^{d}_2 v^*_2.  \label{mass2a}
 \eea
 Complex values of $v_1$ and $v_2$ in general make $M^{u,d}$ complex. So there is a possibility
 to get the flavor non-diagonal phase in the CKM matrix from this setup~\cite{TDLee2, Chen et al}.
 However, it is complicated to show that this can be achieved in general cases.
 In the following, we will use an explicit example to show that this can be achieved.
 
An explicit example of spontaneous generation of the flavor non-diagonal CP violating
phase in the CKM matrix can be given by couplings as follows
 \bea
  Y^u_1=\begin{pmatrix} 1 & 0 & 0 \cr 0 & 0 & 0 \cr 0 & 0 & 0\end{pmatrix}
 K^\dagger_{23} ~g^u, ~~Y^u_2=\begin{pmatrix} 0 & 0 & 0 \cr 0 & 1 & 0 \cr 0 & 0 & 1 \end{pmatrix}
 K^\dagger_{23} ~h^u, \label{mixing1}
 \eea
 and
 \bea
 Y^d_1=K_{13}\begin{pmatrix} 1 & 0 & 0 \cr 0 & 0 & 0 \cr 0 & 0 & 0 \end{pmatrix} K_{12} ~g^d,~~
  Y^d_2=K_{13}\begin{pmatrix} 0 & 0 & 0 \cr 0 & 1 & 0 \cr 0 & 0 & 1 \end{pmatrix} K_{12} ~h^d, \label{mixing2}
 \eea
 where $g^{u,d}$ and $h^{u,d}$ are diagonal matrices with real eigenvalues and they are taken as 
 \bea
 g^{u,d}=y^{u,d} ~v/|v_1|, ~h^{u,d}=y^{u,d} ~v/|v_2| \label{yukawa}
 \eea
 where $y^u$ and $y^d$ are diagonalized real Yukawa couplings appearing in the SM.
 $K_{12,13,23}$ are standard rotation matrices in the standard
 parametrization of the CKM matrix with rotation angles appearing in 1-2, 1-3 or 2-3 entries.
  In the following, we will take $s_{ij}$ and  $c_{ij}$ as the
 sine and cosine of the rotation angle $\theta_{ij}$ in the matrix $K_{ij}$.  We note that  Yukawa couplings
 in (\ref{mixing1}) and (\ref{mixing2}) can be re-defined subject to rotations as $Y^u_i \to O_L Y^u_i O_u$ and 
 $Y^d_i \to O_L Y^d_i O_d$ where $O_{L,u,d}$ are real rotation matrices, and results
 presented in this article are kept intact under this re-definition.
 
 We can see in (\ref{mixing1}), (\ref{mixing2}) and (\ref{yukawa}) that the strength of the Yukawa couplings
 can be much larger than that in the SM. In particular, if $|v_1| \ll v$, e.g. $|v_1| \sim 0.1 v$,
 strength of $Y_1$ can be ten times larger than in the SM, while the strength of $Y_2$, the couplings
 with the second and third generation of quarks would remain almost the same as in the SM.
 Measurements of the Yukawa couplings of the third generation quark, which are so far consistent
 with the SM prediction~\cite{RPP}, would put a constraint on this model. This constraint says that $|v_2|$ should
 be much larger than $|v_1|$. So we can conclude in this model that the strength of the Yukawa couplings
 of the first generation quark should be much larger than that in the SM.
 
The CP phases in $v_1$ and $v_2$ are subject to re-phasing of scalar fields and can be taken in $v_1$:
 \bea
 v_1=|v_1| e^{-i\delta},~~v_2=|v_2|. 
 \eea
 So we can find that
 \bea
 M^u=V_L^u m^u,~~m^u=y^u v, \label{mass11}
 \eea
  and
 \bea
 V^u_L=
 \begin{pmatrix} e^{- i\delta } & 0 & 0 \cr 0 & 1 & 0 \cr 0 & 0 & 1 \end{pmatrix} K^\dagger_{23} \label{mass11-a}
 \eea
 for up-type quarks, and 
 \bea
 M^d=V_L^d m^d,~~m^d=y^d v, \label{mass12}
 \eea
 and 
 \bea
 ~V^d_L=K_{13} \begin{pmatrix} e^{i\delta } & 0 & 0 \cr 0 & 1 & 0 \cr 0 & 0 & 1 \end{pmatrix}  K_{12},
  \label{mass12-a}
 \eea
for down-type quarks. In writing out (\ref{mass11}) and (\ref{mass12}),
(\ref{yukawa}) has been used.

So we can get the CKM matrix
\bea
K=V^{u\dagger}_L V^d_L=K_{23} \begin{pmatrix} e^{i\delta } & 0 & 0 \cr 0 & 1 & 0 \cr 0 & 0 & 1 \end{pmatrix}
K_{13} \begin{pmatrix} e^{i\delta } & 0 & 0 \cr 0 & 1 & 0 \cr 0 & 0 & 1 \end{pmatrix}  K_{12}. 
\eea
It is equivalent to the standard parametrization of the CKM matrix~\cite{CK, RPP} 
via a vector-like transformation of up quark which does not change $\theta$. That is, with
\bea
u_L \to e^{2 i\delta} u_L, u_R \to e^{2i\delta} u_R \label{transformation1}
\eea
we  can get the standard parametrization of CKM matrix
\bea
K\to K=K_{23} \begin{pmatrix} e^{-i\delta } & 0 & 0 \cr 0 & 1 & 0 \cr 0 & 0 & 1 \end{pmatrix}
K_{13} \begin{pmatrix} e^{i\delta } & 0 & 0 \cr 0 & 1 & 0 \cr 0 & 0 & 1 \end{pmatrix}  K_{12}.  \label{CKM}
\eea

We conclude that spontaneous generation of the CP violating phase in the CKM matrix
can be achieved in model described in this article. We use an explicit example to show this possibility. 
In particular, we show that spontaneous generation of the CP violating phase in the CKM
matrix can be achieved using matrices of Yukawa couplings with rank less than three.

 \section{\large FCNC and strong $\theta$ term with spontaneous generation of CP violation}\label{sec3}
 In this section we show how FCNC can be suppressed in the model presented in the present article.
 We then show how strong CP problem is avoided.

 One of the main problem associated with (\ref{mass2}) is the
 that unitary transformations that diagonalize the mass matrix (\ref{mass2}) do not necessarily diagonalize the
 Higgs couplings to quarks in (\ref{yukawa2}).  So tree level FCNC could be present and  this type of theories
 of spontaneous generation of CP violation may encounter difficulty in this aspect, as pointed out in \cite{Geng}
 for model presented in \cite{TDLee2}.
  
 We show how dangerous FCNC can be avoided. Taking
\bea
M^{u,d}=M^{u,d}_1+M^{u,d}_2\label{mass3}
\eea
where $M^u_{1,2}= Y^u_{1,2} v_{1,2}$ and $M^d_{1,2}= Y^d_{1,2} v^*_{1,2}$, 
we assume $M^{u,d}_2$ is the dominant contribution to $M^{u,d}$, i.e.
\bea
||M^{u,d}_1|| \ll |M^{u,d}_2||. \label{mass3b}
\eea
Note that $M_1$ or $M_2$ is proportional to real matrix $Y_1$ or $Y_2$ and 
diagonalizing $M_1$ or $M_2$ does not need complex matrices. 
So  diagonalizing $M_1$ or $M_2$ does not change strong $\theta$ and we can work in a base that one of $M_1$ and $M_2$ is diagonalized.
In this base non-zero elements of $M^{u,d}_1$ can be taken at most at order of $m_{u,d}$ with $m_{u,d}$ being the
masses of up and down quarks.

Taking $M^{u,d}_2$ as rank two and working in the base that $M^u_2$ is diagonalized
\bea
M^u_2=\textrm{diag}\{0, x_2,x_3 \}\label{mass4}
\eea
we can write 
\bea
M^u=\begin{pmatrix} x_{11} & x_{12} & x_{13} \cr
x_{21} & x_{22} +x_2 & x_{23}  \cr x_{31} & x_{32} & x_{33} +x_3 \end{pmatrix}, \label{mass5}
\eea
where $x_{ij}$ comes from $M^u_1$. In this form of mass matrix, (\ref{mass3b}) means $|x_{ij}|\ll x_{2,3}$ for $M^u_1$ and $M^u_2$.
In particular, non-zero $x_{ij}$  would be at most of order $m_u$. 
$x_2\approx m_c$ and $x_3 \approx m_t$ with $m_{c,t}$ being the masses of charm and top quarks.

If $x_{ij}=0$ except $x_{11}$,  (\ref{mass5}) is already diagonalized and
no extra flavor mixing is needed. In this case, $M_1^u$ and $M_2^u$ can be diagonalized simultaneously.  
In general, off-diagonal matrix element $x_{ij}$ may not be zero, and $M_1^u$ and $M_2^u$  can not be diagonalized simultaneously.
So tree level FCNC can be present.
However,  for $|x_{ij}| \ll x_{2,3}$ the matrices $V_L$ and $V_R$ that further diagonalize $M^u$ in (\ref{mass5})
are all close to unit matrix. More specifically, we can write $M^u=V_L m^u V_R^\dagger$ where $m^u$ is the diagonalized
mass matrix of up-type quarks with real eigenvalues $m_{1,2,3}$ and possible $U(1)$ factors, which do not change the conclusion about FCNC,
have been suppressed.  To first order we find
\bea
V_L\approx \begin{pmatrix} 1 &  a_{12} & a_{13} \cr -a_{12}^* & 1 & a_{23} \cr -a_{13}^* & -a_{23}^* & 1\end{pmatrix},~
V_R\approx \begin{pmatrix} 1 &  b_{12} & b_{13} \cr -b_{12}^* & 1 & b_{23} \cr -b_{13}^* & -b_{23}^* & 1\end{pmatrix},
\eea
where $a_{ij}$ and $b_{ij}$  satisfy $|a_{ij}|\ll 1$ and $|b_{ij}|\ll 1$.   $a_{ij}$ and $b_{ij}$ are found to be 
\bea
a_{ij}=\frac{x_{ij} m_j+x_{ji}^* m_i}{m_j^2-m_i^2},
~b_{ij}=\frac{x_{ij} m_i+x_{ji}^* m_j}{m_j^2-m_i^2}, ~\textrm{for $ i<j$}.
\eea
For eigenvalues in $m^u$, we have $m_{2,3}\approx x_{2,3}$ and $m_1\approx x_{11}$.
Since $m_j \gg m_i$ for $j >i$, we can find $a_{ij}\approx x_{ij}/m_j$ and $b_{ij}\approx x_{ji}^*/m_j$.
We can see that non-zero off-diagonal elements in $V_{L,R}$ are either of order $m_u/m_c$ or of order $m_u/m_t$.

After diagonalizing the mass matrix $M^u$, the coupling of $\phi_1$ with up-type quarks, which originally mixes flavors,
still mixes flavors. The strength of this FCNC coupling is $x_{ij}/|v_1|\sim m_u/|v_1|$. 
After diagonalizing the mass matrix $M^u$, the coupling of $\phi_2$ with up-type quarks, which is originally flavor diagonal,
gives rise to new FCNC couplings. For example, a flavor diagonal coupling $(x_i/v_2) {\bar u}_L^i u^i_R \phi_2^0$, with $\phi_2^0$
being the neutral component of $\phi_2$, becomes $  (x_i/v_2) (V_L)_{ij}^* (V_R)_{ik}{\bar u}^j_L u^k_R \phi_2^0 $ after
field re-definition using $V_{L,R}$.
At first order, it gives rise to FCNC couplings  $  (x_i/v_2) (V_L)_{ij}^* {\bar u}^j_L u^i_R \phi_2^0 $ and
 $  (x_i/v_2)  (V_R)_{ij}{\bar u}^i_L u^j_R \phi_2^0 $  for $j\neq i$.
 As shown above, the off-diagonal matrix elements $(V_L)_{ij}$ and $(V_R)_{ij}$ with $i\neq j$ all
 have a strength $\sim |x_{ij}|/max(m_i,m_j)$ or $\sim |x_{ji}|/max(m_i,m_j)$ . 
 So we can find that these FCNC couplings induced in couplings with $\phi_2$ have 
 strength $\sim x_i/v_2 \times (|x_{ij}|  ~\textrm{or} ~|x_{ij}|)/max(m_i,m_j) \lsim m_u/v_2$.
 Summarizing these two cases, the strengths of the FCNC couplings with up-type quarks are suppressed to be $\lsim m_u/|v_1|$.

Similarly, one can show that possible FCNC interactions of Higgs with down-type quarks are also suppressed
to be less than order $m_d/|v_1|$ if taking $M^d_2$ as the dominant contribution to $M^d$.
If magnitude of $v_1$ is not extremely small, FCNC couplings given by Yukawa couplings in (\ref{yukawa2})
can be safely neglected. For example, if $|v_1|\sim 0.1 \times v$, FCNC Higgs coupling with up-type quarks would
be at most at order $10^{-4}$.  Any possible FCNC processes induced by these couplings would be suppressed
by the square of this FCNC amplitude, i.e. suppressed by a factor of order $10^{-8}$.
However, if the magnitude of $v_1$ is extremely small, e. g. $|v_1| \sim 10^{-3} ~v$,  the magnitude of $Y_1$ would be
large and there could be dangerous FCNC couplings arising from it. To avoid possibly large FCNC couplings a hierarchy between
$Y_1$ and $Y_2$ is preferred. This implies that $|v_1|$ should not be very small. We assume $|v_1| > 0.01 ~v$.

We note that our arguments for suppressing FCNC are based on the assumption that a hierarchy can exist in $Y_1$ and $Y_2$. 
For this assumption to hold, radiative corrections should not change the hierarchy.
In fact, the radiative correction to $Y_1$ from couplings of $\phi_2$  will be proportional to elements of $Y_1$, so that
the hierarchy between $Y_1$ and $Y_2$ is not affected by the radiative corrections.
This means that the suppression of FCNC is robust against quantum correction.
This is because we have taken $Y_2^{u,d}$ as rank two and if setting $Y^{u,d}_1=0$ there is a chiral symmetry 
appearing in the Lagrangian which protects the hierarchy.

Now we come to explain that the strong $\theta$ is naturally zero in this model. Using (\ref{mass2}) one can easily show that
the determinant $\det(M^u M^d)$ is real and 
zero correction to $\theta$ can be achieved if taking the rank of $Y_1$ and $Y_2$ both less than 3.
For example, if taking $Y_1$ or $M_1$ as rank one and $Y_2$ or $M_2$ as rank two
we can write them as
\bea
M_1&&=\begin{pmatrix}  x^u_{11} v_1 & x^u_{12} v_1 & x^u_{13} v_1 \cr
a_u x^u_{11} v_1 & a_u x^u_{12} v_1 & a_u x^u_{13} v_1 \cr
b_u x^u_{11} v_1 & b_u x^u_{12} v_1 & b_u x^u_{13} v_1 
\end{pmatrix}, \label{mass6} \\
M_2 &&=\textrm{diag} \{ 0,  y^u_2 v_2, y^u_3 v_2 \} \label{mass7},
\eea
where $x^u_{ij}$, $y^u_i$, $a_u$ and $b_u$ are all real numbers and $y^u_{2,3}$ are eigenvalues of $Y^u_2$.
We can find 
\bea
det(M^u)=det(M^u_1+M^u_2)=x^u_{11} y^u_2 y^u_3 v_1 v_2 v_2 \label{mass8}
\eea
Similar expression holds for $M^d$
\bea
M^d= Y^d_1 v_1^* +Y^d_2 v_2^* .
\eea
and we get 
\bea
det(M^d)=x^d_{11} y^d_2 y^d_3 v_1^* v_2^* v_2^* . \label{mass9}
\eea
So we get
\bea
det(M^u M^d)=x^u_{11} y^u_2 y^u_3 x^d_{11} y^d_2 y^d_3 |v_1|^2 |v_2|^4. \label{mass10}
\eea
(\ref{mass10}) is real and arg$(det(M^u M^d))=0$. So the correction to $\theta$ is zero as can be seen in
(\ref{strongtheta2}).  This statement relies on the fact that $Y_1$ and $Y_2$ all have ranks
less than three, and in particular the sum of the ranks of $Y_1$ and $Y_2$ equals to three.
Radiative corrections can change this feature and give rise to nonzero correction
to $\theta$.

We conclude that in model presented in this article 
the QCD $\theta$ is zero at tree level after spontaneous generation of the CP violating phase in the CKM matrix
and FCNC can be naturally suppressed.   In model presented here
we have assumed that the two matrices of Yukawa couplings both have rank less than three,
and the sum of the ranks of two matrices of Yukawa couplings equals to three.
So the two Higgs doublets complement to each other in the sense that they together
give rise to the complete quark mass matrices and none of them is enough without the help of other Higgs doublet.
In this sense, we can call this model of two Higgs doublets as Complementary two Higgs doublet model(C2HDM).

As discussed in the previous section,  spontaneous generation of the CP violating phase in the CKM
matrix can be achieved using matrices of Yukawa couplings with rank less than three and
in particular in C2HDM.
Moreover, the Yukawa couplings used in the last section, (\ref{mixing1}) and (\ref{mixing2}),
satisfy the assumption in this section and indeed lead to zero contribution to $\theta$ at tree level.

Since the discussion on strong $\theta$ in this section depends on the assumption of Yukawa couplings,
it's natural to ask what is the effect of radiative correction on the assumed Yukawa couplings
and what is the effect on the size of the induced strong $\theta$. In the next section we will study this question.

\section{\large Radiative correction to Yukawa coupling and strong $\theta$}\label{sec4}
In this section we study radiative correction to Yukawa couplings and the correction
to $\theta$ term arising from it. A general discussion on the correction to $\theta$ term seems 
very complicated. We are not going to do a general discussion, but rather to show that the radiative correction to
 (\ref{mixing1}) and (\ref{mixing2})  would lead to a correction to strong $\theta$
at order $10^{-12}\sim 10^{-10}$ at one-loop level. With this example, we illustrate that a small enough strong $\theta$, being
consistent with experimental bound, can be achieved in model of spontaneous generation of 
the CP violating phase in the CKM matrix.

One-loop radiative correction to Yukawa couplings can be read out in their Renormalization Group Equation(RGE)
as shown in (\ref{RGE1}) and (\ref{RGE2}).  Yukawa couplings in  (\ref{mixing1}) and (\ref{mixing2})
have a nice feature
\bea
\textrm{Tr} [ Y^u_i Y^{u\dagger}_j]=\textrm{Tr} [ Y^d_i Y^{d\dagger}_j]=0, \textrm{for $i\neq j$ }. \label{mixing3}
\eea
If we further assume $Y^l_i$, the Yukawa coupling of charged leptons, also has a similar feature
\bea
\textrm{Tr} [Y^{l\dagger}_i Y^l_j ]=0,  \textrm{for $i\neq j$}, \label{mixing4}
\eea
RGEs in (\ref{RGE1}) and (\ref{RGE2}) can be simplified. In particular, the second term in (\ref{RGE1}) or (\ref{RGE2}) can be combined with
the first term and part of the last term can be combined with the third term. Moreover, in this case $Y^l_i$ would appear in factor $A^{u,d}_i$
and can be omitted in future discussion.  So we arrive at (\ref{RGE3}) and (\ref{RGE4}) .

Using  (\ref{RGE3}) and (\ref{RGE4}) we can see that the one loop corrected Yukawa couplings are
\bea
Y^{'u}_1&&=Y^u_1+\epsilon [-A^u_1 Y^u_1 +B^u_1 Y^u_1 +Y^u_1 C^u-2 Y^d_2 Y^{d\dagger}_1 Y^u_2], \label{yukawa-a1}\\
Y^{'u}_2&&=Y^u_2+\epsilon [-A^u_2 Y^u_2 +B^u_2 Y^u_2 +Y^u_2 C^u], \label{yukawa-a2}\\
Y^{'d}_1&&=Y^d_1+\epsilon [-A^d_1 Y^d_1 +B^d_1 Y^u_1 +Y^d_1 C^d], \label{yukawa-a3}\\
Y^{'d}_2&&=Y^d_2+\epsilon [-A^d_2 Y^d_2 +B^d_2 Y^d_2 +Y^d_2 C^d], \label{yukawa-a4}
\eea
where we have used (\ref{result2}) and $\epsilon=log(\mu/\Lambda)/(16\pi^2)$.   $\epsilon$ is a small number and
$\epsilon^2$ would be smaller than around $10^{-3}$ for $\Lambda$ lower than around $10^5$ GeV
which means the new physics scale is no more than three orders of magnitude higher than the electroweak scale.
$Y^d_2 Y^{d\dagger}_1 Y^u_2$ in (\ref{yukawa-a1}) is found to be rank one  and is given in (\ref{result1}). 
Other $Y^d_i Y^{d\dagger}_j Y^u_i$ and $Y^u_i Y^{u\dagger}_j Y^d_i$ terms in (\ref{RGE3}) and (\ref{RGE4}) 
are found to be zero.
The mass matrices of up-type and down-type quarks are obtained as
\bea
M^{'u} &&=Y^{'u}_1 v_1 +Y^{'u}_2 v_2 \nnb \\
&& =X^u_1 Y^u_1 v_1+ \epsilon Y^u_1 C^u v_1+
X^u_2 Y^u_2 v_2+ \epsilon Y^u_2 C^u v_2 -2 \epsilon Y^d_2 Y^{d\dagger}_1 Y^u_2 v_1, \label{Mmatrix1} \\
M^{'d} &&=Y^{'d}_1 v^*_1 +Y^{'d}_2 v^*_2\nnb \\
&& =X^d_1 Y^d_1 v^*_1+ \epsilon Y^d_1 C^d v^*_1+X^d_2 Y^d_2 v^*_2+ \epsilon Y^d_2 C^d v^*_2,\label{Mmatrix2}
\eea
where $X^{u,d}_i=1-\epsilon A^{u,d}_i+\epsilon B^{u,d}_i$($i=1,2$) are real matrices with rank three.

Now we can compute the determinant of $M^{'u}$ and $M^{'d}$. 
The leading term of $\det(M^{'u})$  is proportional to $v_1 v^2_2$  and is $v_1 v_2^2 g^u_1 h^u_2 h^u_3$. 
Possible corrections to the $v_1 v^2_2$ term do not change the conclusion of the discussion below and will be omitted.
Sub-leading terms in $\det(M^{'u})$ can be proportional to $v^3_1$, $v_1^2 v_2$ or $v^3_2$.
The term proportional to $v_2^3$ comes from $\det(v_2 X^u_2 Y^u_2+\epsilon v_2 Y^u_2 C^u)$
and as shown in Appendix B  it is zero. Similarly, one can show that the term proportional to $v_1^3$ vanishes.
The leading non-zero correction is the term proportional to $v_1^2 v_2$ as given in (\ref{v1square-v2-2}) and   (\ref{v1square-v2-3}) .
Thus, we obtain $\det (M^{'u})$ as
\bea
&& \det (M^{'u})=v_1 v_2^2 g^u_1 h^u_2 h^u_3 \nnb \\
&&~~~-3 \epsilon^2 v_1^2 v_2 (g^d_2 h^d_2-g^d_1 h^d_1)s_{13}c_{12} s_{12}c_{23}s_{23}
[(h^u_3)^2-(h^u_2)^2] g^u_1 h^u_2 h^u_3,\label{determinant1}
\eea
where smaller correction in (\ref{v1square-v2-3}) has been neglected.

The leading term of $\det(M^{'d})$  is proportional to $v_1^* (v^*_2)^2$ and 
is $v^*_1 (v^*_2)^2 g^d_1 h^d_2 h^d_3$ . Possible corrections to the
 $v_1^* (v^*_2)^2$ term do not change the conclusion of the discussion below and will be omitted.
 Sub-leading terms in $\det(M^{'d})$ can be proportional to $(v_1^*)^3$, $(v_1^*)^2 v_2^*$ or $(v^*_2)^3$.
The term proportional to $(v_1^*)^3$ comes from $\det(v^*_1 X^d_1 Y^d_1+\epsilon v^*_1 Y^d_1 C^d)$
and as shown in Appendix B it is zero.  Terms proportional to $(v^*_2)^3$ and $(v^*_1)^2 v^*_2$ 
are calculated in (\ref{v1square-v2-4}) and (\ref{v2-cubic-2}). Combining these results we can get
\bea
&&\det(M^{'d})=v^*_1 (v^*_2)^2 g^d_1 h^d_2 h^d_3 \nnb \\
 &&~~~+\frac{1}{2} \epsilon^2  [(v^*_2)^3 +(v^*_1)^2 v^*_2] s_{13} c_{12} s_{12} c_{23} s_{23} g^d_1 h^d_2 h^d_3 [(h^u_3)^2-(h^u_2)^2]
 (g^d_2 h^d_2 -g^d_1 h^d_1). \label{determinant2}
\eea
A common factor in  (\ref{determinant1}) and (\ref{determinant2}) is $ s_{13} c_{12} s_{12} c_{23} s_{23}  (g^d_2 h^d_2 -g^d_1 h^d_1)
\sim 10^{-5} (m_s/v)^2 (v^2/|v_1 v_2|) \sim 10^{-11} \times (v^2/|v_1 v_2|)$. One can see that for $|v_1| \ll |v_2|$ correction in
 (\ref{determinant1}) is of order $10^{-11} \epsilon^2$ and correction in (\ref{determinant2}) is of order $10^{-11} (v^2/|v_1|^2) \epsilon^2 $.
 So we have 
\bea
\det( M^{'u} M^{'d})=|v_1|^2 |v_2|^4 g^u_1 h^u_2 h^u_3  g^d_1 h^d_2 h^d_3 \times [1+{\cal O} (10^{-11}) \times  {v^2 \over |v_1|^2}\times\epsilon^2 ]
\eea
We can see that the radiative correction to the determinant of $M^u M^d$ is of order $10^{-12}$
for $|v_1| \sim 0.1 v$ and for $\epsilon^2\approx  10^{-3}$  which corresponds to the
new physics scale being three orders of magnitude higher than the electroweak scale.
The radiative correction would be smaller if
new physics scale is closer to the electroweak scale.  
One can also see that if $|v_1|$ is too small, e.g. $|v_1|\lsim 10^{-2}$ v, the radiative correction
would be too large and it could give rise to a strong $\theta$ reaching the experimental bound. 
Since the Yukawa couplings of the first generation fermions are proportional to $1/v_1$,
$|v_1|$ should not be too small as argued for naturally suppressing possible FCNC couplings.
In particular we have assumed that $|v_1| > 0.01$ v.
We can conclude that for reasonable values of parameters, 
the radiative correction to the determinant of $M^u M^d$ is smaller than order of $10^{-12}\sim 10^{-10}$. 
So its correction to strong $\theta$ is also smaller than order $10^{-12}\sim10^{-10}$.

Combining the conclusions in the last section and this section, we can see that in the scenario discussed in this article,
i.e. with (\ref{mixing1}) and (\ref{mixing2}), strong $\theta$ is zero at leading order and can be generated at one-loop
level, but is smaller than around $10^{-12}\sim 10^{-10}$. This prediction can be tested in future EDM experiment~\cite{futureEDM}.

 \section{\large Conclusion}\label{sec5}
In summary, we have presented a model of spontaneous CP violation.
CP symmetry is exact before the spontaneous symmetry breaking. 
The CP violating phase in the CKM matrix is generated spontaneously in this model.
We show that it is possible to achieve a zero strong $\theta$ even after the
spontaneous breaking of CP symmetry in such kind of model.

We show that zero strong $\theta$ term can be achieved if the two Higgs doublets involved in the setup
are complementary in the sense that they are both needed to describe the quark masses
and none of them is enough. To be specific, the ranks of the two Yukawa couplings with a specific type of quark, 
say up-type quark or down-type quark, can be rank two and rank one and their sum is three. 
We have called this kind of model of two Higgs doublets as  C2HMD, the Complementary 2HDM.  
It's straightforward to show that similar conclusion can be achieved if there are three Higgs doublets
and each them couple to the quark fields with a rank one Yukawa coupling, similar to the case that
each Higgs doublet coupled with one generation of quarks.

In a specific model with specific Yukawa couplings, we have studied the radiative correction to the assumed Yukawa couplings
and have discussed the robustness of the above statement on strong $\theta$. We find in this example that
correction to strong $\theta$ can vary from $10^{-12}$, much smaller than the experimental bound, to $10^{-10}$ reaching the experimental bound,
depending on the Yukawa couplings of the first generation quarks.
Using this example, we demonstrate that it is possible to have a very small strong $\theta$ term
in model of spontaneous generation of the CP violating phase in the CKM matrix
even if radiative correction to the assumed scenario is considered into account.
A general discussion on this part seems complicated and we have left it to future study.

We have shown that in the set-up discussed in the present article, say C2HDM,  not only the CP violating phase
in CKM matrix can be generated spontaneously and strong $\theta$ is naturally zero or very small,
but also the the dangerous FCNC can be naturally suppressed. The point is that one of the Higgs doublets
in the complementary pair of the two Higgs doublets can be the dominant one and FCNC
would naturally vanish without the other complementary Higgs doublet. So the appearance of FCNC
coupling would be proportional to the strength of the other Yukawa coupling and is suppressed by
quantities $\sim m_u/|v_1|$ or $m_d/|v_1|$.

We note that one interesting consequence of the model is that the
Yukawa coupling of the first generation is around $\sim m_q/|v_1|$ which can
be much larger than the corresponding Yukawa couplings in the SM.  
For example, the Yukawa couplings of the first generation quarks can be ten times larger than that in the SM if $|v_1| \sim 0.1 v$, or even
larger if $|v_1|$ is even smaller. The strength of the coupling of the light neutral Higgs with first generation quarks would be
proportional to $\sim \sin\alpha ~m_q/|v_1|$ with $\alpha$ being the mixing angle of neutral Higgs field. The strength
of the coupling of the heavy neutral Higgs would be proportional to $\sim \cos\alpha ~m_q/|v_1|$. They both have
a possibility to be much larger than what usually expected in 2HDMs. This may give rise to interesting implications for Higgs phenomenology.
Since the radiative correction to strong $\theta$ also depends on the Yukawa couplings of the first generation quarks, 
measuring and testing the Yukawa couplings seem to be a very interesting subject to study.

We note that the SM can not give a successful explanation of the baryon-number generation in the universe. Physics beyond the
SM and new source of CP violation are needed to implement a baryon-number generation in the early universe.
Since CP symmetry is broken spontaneously  in our model, sufficient baryon-number generation should also be implemented by the model.
Some other interesting topics include the CP violating phase in the leptonic sector and the impact to neutrino mixings~\cite{others}.
In the present paper we do not discuss all these related issues and leave them to future research.

\begin{center} {\bf Acknowledgements} \end{center}
 This work is supported by National Science Foundation of
 China(NSFC), grant No. 11375065,  and Shanghai Key Laboratory
 of Particle Physics and Cosmology, grant No. 15DZ2272100. 

\section*{Appendix A}
RGEs for Yukawa coupling $Y^u_i$ and $Y^d_i$ for a general two-Higgs doublet model are ~\cite{FLS}:
\bea
16\pi^2 \frac{d}{dln\mu} Y^u_i &&=-A^u Y^u_i+\sum_j \textrm{Tr}[ N_c(Y^u_iY^{u\dagger}_j+Y^d_j Y^{d\dagger}_i)
+Y^{l\dagger}_i Y^l_j] Y^u_j  \nnb \\
&& +\frac{1}{2}\sum_j(Y^u_j Y^{u\dagger}_j+Y^d_j Y^{d\dagger}_j) Y^u_i
+Y^u_i \sum_j Y^{u\dagger}_j Y^u_j-2 \sum_j Y^d_j Y^{d\dagger}_i Y^u_j, \label{RGE1} \\
16\pi^2 \frac{d}{dln\mu} Y^d_i &&=-A^d Y^d_i+\sum_j \textrm{Tr} [ N_c(Y^d_iY^{d\dagger}_j+Y^u_j Y^{u\dagger}_i)
+Y^l_i Y^{l\dagger}_j ] Y^d_j  \nnb \\
&& +\frac{1}{2}\sum_j(Y^u_j Y^{u\dagger}_j+Y^d_j Y^{d\dagger}_j) Y^d_i
+Y^d_i \sum_j Y^{d\dagger}_j Y^d_j-2 \sum_j Y^u_j Y^{u\dagger}_i Y^d_j,\label{RGE2}
\eea
where $Y^l_i$ is the Yukawa coupling of charged lepton, $N_c=3$, and
\bea
A^u=8 g^2_3 +\frac{9}{4} g^2_2+\frac{17}{12} g^2_1, 
A^d=8 g^2_3 +\frac{9}{4} g^2_2+\frac{5}{12} g^2_1
\eea
with $g_{1,2,3}$ the gauge couplings of $U(1)_Y$, $SU(2)_L$ and $SU(3)_C$ groups respectively.

With assumption of (\ref{mixing3}) and (\ref{mixing4}),   (\ref{RGE1}) and (\ref{RGE4}) can be written as
\bea
16\pi^2 \frac{d}{dln\mu} Y^u_i &&=-A^u_i Y^u_i+ B^u_i Y^u_i +Y^u_i C^u-2 \sum_{j\neq i} Y^d_j Y^{d\dagger}_i Y^u_j, \label{RGE3}\\
16\pi^2 \frac{d}{dln\mu} Y^d_i &&=-A^d_i Y^d_i+B^d_i Y^d_i + Y^d_i C^d-2 \sum_{j\neq i} Y^u_j Y^{u\dagger}_i Y^d_j, \label{RGE4}
\eea
where using (\ref{mixing1}) and (\ref{mixing2}) $A^{u,d}_i$, $B^{u,d}_i$ and $C^{u,d}_i$ are given as 
\bea
A^{u,d}_i &&=A^{u,d}-\textrm{Tr}[ N_c (Y^u_i Y^{u\dagger}_i+Y^d_i Y^{d\dagger}_i)+Y^l_i Y^{l\dagger}_i ], \label{Aud} \\
B^u_1&&= B-2 K_{13} ~\textrm{diag}\{ g^d_1)^2 c^2_{12}+ (g^d_2)^2 s^2_{12},0, 0 \} ~K^\dagger_{13}, \label{Bud1} \\
B^u_2 &&=B-2K_{13}  ~\textrm{diag}\{ 0, (h^d_1)^2 s^2_{12}+ (h^d_2)^2 c^2_{12},(h^d_3)^2 \} ~K_{13}^\dagger, \label{Bud2} \\
B^d_1 &&=B-2 K^\dagger_{23} ~\textrm{diag}\{(g^u_1)^2,  0, 0 \}~K_{23}, \label{Bud3} \\
B^d_2 &&=B-2 K^\dagger_{23} ~\textrm{diag}\{0,  (h^u_2)^2, (h^u_3)^2 \} ~K_{23} \label{Bud4},
\eea
where 
\bea
B&&=\frac{1}{2}K^\dagger_{23} ~\textrm{diag}\{(g^u_1)^2, (h^u_2)^2,(h^u_3)^2 \} ~K_{23} \nnb \\
&&+ \frac{1}{2} K_{13} ~\textrm{diag}\{(g^d_1)^2 c^2_{12}+ (g^d_2)^2 s^2_{12}, (h^d_1)^2 s^2_{12}+ (h^d_2)^2 c^2_{12},(h^d_3)^2 \} 
~K_{13}^\dagger, 
\label{Bud5}
\eea
and 
\bea
C^u &&= \textrm{diag} ~\{(g^u_1)^2, (h^u_2)^2,(h^u_3)^2 \}, \label{Cu1} \\
C^d &&= \begin{pmatrix}   (g^d_1)^2 c^2_{12}+(h^d_1)^2 s^2_{12} & (g^d_1 g^d_2-h^d_1 h^d_2) s_{12} c_{12} & 0 \cr
 (g^d_1 g^d_2-h^d_1 h^d_2) s_{12} c_{12} & (g^d_2)^2 s^2_{12}+(h^d_2)^2 c^2_{12} & 0 \cr
 0 & 0 & (h^d_3)^2    \end{pmatrix}. \label{Cud2}
\eea

\section*{Appendix B}
Using  (\ref{mixing1}) and (\ref{mixing2}) one can find that the last terms in (\ref{RGE3}) and (\ref{RGE4}) are
\bea
Y^d_2 Y^{d\dagger}_1 Y^u_2=\begin{pmatrix}0 & 0 & 0 \cr 0 & 0 & -s_d \cr 0 & 0 & 0 \end{pmatrix}
K^\dagger_{23} h^u, \label{result1}
\eea
where $s_d=s_{13}c_{12}s_{12}(h^d_2 g^d_2-h^d_1 g^d_1)$, and 
\bea
Y^d_1 Y^{d\dagger}_2 Y^u_1=Y^u_1 Y^{u\dagger}_2 Y^d_1=Y^u_2 Y^{u\dagger}_1 Y^d_2=0. \label{result2}
\eea
One can also find from $(Y^u_2)_{1i}=(Y^u_2)_{i1}=0$ that
\bea
(Y^u_2 C^u)_{1i}=(Y^u_2 C^u)_{i1}=(X^u_2 Y^u_2)_{i1}=0, ~i=1,2,3. \label{result3}
\eea
\vskip 0.5cm

Determinant of $M^{'u}$ in (\ref{Mmatrix1}) is calculated as follows.\\
1) $v^3_2$ term in $\det(M^{'u})$ comes from $\det(X^u_2 Y^u_2 v_2+\epsilon Y^u_2 C^u v_2)$.
Using the fact that $Y^u_2$ is rank two and $(Y^u_2 C^u)_{1i}=0$  we can get
\bea
&&\det(X^u_2 Y^u_2 v_2+\epsilon Y^u_2 C^u v_2)  = v_2^3 \varepsilon^{ijk}
[ \epsilon (X^u_2 Y^u_2)_{1i} (X^u_2 Y^u_2)_{2j} (Y^u C^u)_{3k} \nnb \\
&& ~~~~~~+\epsilon (X^u_2 Y^u_2)_{1i}  (Y^u C^u)_{2j} (X^u_2 Y^u_2)_{3k} 
+\epsilon^2 (X^u_2 Y^u_2)_{1i}  (Y^u C^u)_{2j}   (Y^u C^u)_{3k} ]. \label{v2-cubic}
\eea
Since one of the $i,j,k$ has to be 1, according to (\ref{result3}), none of the
three terms in (\ref{v2-cubic}) is nonzero. So the term proportional to $v_2^3$ is zero.

2) $v_1^3$ terms in  $\det(M^{'u})$ comes from $\det(X^u Y^u_1 v_1 +\epsilon Y^u_1 C^u v_1 -2 \epsilon Y^d_2 Y^{d\dagger}_1 Y^u_2 v_1)$.
Since $(Y^d_2 Y^{d\dagger}_1 Y^u_2)_{1i}=(Y^d_2 Y^{d\dagger}_1 Y^u_2)_{3i}=0$ as shown in (\ref{result1}),
we can get
\bea
&& \det[X^u_1 Y^u_1 v_1 +\epsilon Y^u_1 C^u v_1 -2 \epsilon Y^d_2 Y^{d\dagger}_1 Y^u_2 v_1] \nnb \\
&&~~~~~~=-2 \epsilon^2 v_1^3 \varepsilon^{ijk} [ (X^u_1 Y^u_1)_{1i}  (Y^u_1 C^u)_{3k}
+ (Y^u_1 C^u)_{1i} (X^u_1 Y^u_1)_{3k} ] (Y^d_2 Y^{d\dagger}_1 Y^u_2)_{2j}, \label{v1-cubic}
\eea
Since $ (Y^u_1 C^u)_{3k}=0$,
$(X^u_1 Y^u_1)_{3k}=0$ for $k=2$ or $3$ and $(Y^u_1 C^u)_{1i}=0$ for $i=2$ or $3$, we can see that
 (\ref{v1-cubic}) is zero.
 
 3) $v_1^2 v_2$ term $\det(M^{'u})$ have two factors of $v_1$. Similar to discussion above for (\ref{v1-cubic}),
 we can find that  $\varepsilon^{ijk} (X^u_1 Y^u_1)_{ai} (Y^u_1 C^u)_{bj}=0$ and 
 these two factors of $v_1$ can not both come from $X^u_1 Y^u_1 +\epsilon Y^u_1 C^u$.
 One can find that the $v_1^2 v_2$ term is
 \bea
&& -2\epsilon v_1^2 v_2 \varepsilon^{ijk} [ (X^u_1 Y^u_1 +\epsilon Y^u_1 C^u)_{1i} (Y^d_2 Y^{d\dagger}_1 Y^u_2)_{2j} 
(X^u_2 Y^u_2 +\epsilon Y^u_2 C^u)_{3k}\nnb \\
 && ~~~~~~+ (X^u_2 Y^u_2  +\epsilon Y^u_2 C^u)_{1i} (Y^d_2 Y^{d\dagger}_1 Y^u_2)_{2j} (X^u_1 Y^u_1 +\epsilon Y^u_1 C^u)_{3k}].
 \label{v1square-v2-1}
 \eea
 Using $ (Y^d_2 Y^{d\dagger}_1 Y^u_2)_{2j} =s_d (Y^u_2)_{3j}$ as shown in (\ref{result1}),  
 the first term in (\ref{v1square-v2-1}) can be found to be
 \bea
 &&-2\epsilon v_1^2 v_2 \varepsilon^{ijk}  (X^u_1 Y^u_1 +\epsilon Y^u_1 C^u)_{1i}  s_d (Y^u_2)_{3j} [ (1-\epsilon A^u_2) Y^u_{3k}
 +  (\epsilon B^u_2 Y^u_2 +\epsilon Y^u_2 C^u)_{3k} ] \nnb \\
 &&~~~=-2\epsilon^2v_1^2v_2\varepsilon^{ijk}(X^u_1 Y^u_1 +\epsilon Y^u_1 C^u)_{1i} s_d (Y^u_2)_{3j}(B^u_2 Y^u_2 +\epsilon Y^u_2 C^u)_{3k} \nnb\\
 &&~~~\approx -2\epsilon^2 v_1^2 v_2 s_d \varepsilon^{ijk} (Y^u_1)_{11}\delta_{i1} (Y^u_2)_{3j}
 [ (B^u_2)_{32} (Y^u_2)_{2k}+(Y^u_2)_{3k} (C^u)_{kk} ] \nnb \\
 && ~~~=-2\epsilon^2 v_1^2 v_2 s_d (B^u_2)_{32} (Y^u_1)_{11} [ \varepsilon^{1jk} (Y^u_2)_{3j} (Y^u_2)_{2k} 
 +(Y^u_2)_{32} (Y^u_2)_{33} ((C^u)_{33}-(C^u)_{22}) ]\nnb \\
 &&~~~=-3\epsilon^2 s_d v_1^2 v_2 [ (h^u_3)^2-(h^u_2)^2] g^u_1 h^u_2 h^u_3 c_{23} s_{23}, \label{v1square-v2-2}
  \eea
  where $s_d$ is given after (\ref{result1}).
  The second term in (\ref{v1square-v2-1}) can be found to be
  \bea
  &&-2\epsilon^3 v_1^2 v_2 \varepsilon^{ijk} s_d (B^u_2 Y^u_2  + Y^u_2 C^u)_{1i}(B^u_1 Y^u_1  + Y^u_1 C^u)_{3k} \nnb \\
  &&=-2\epsilon^3 v_1^2 v_2 \varepsilon^{ijk} s_d (B^u_2)_{12} (B^u_1)_{31} (Y^u_2)_{2i} (Y^u_2)_{3j} (Y^u_1)_{11} \delta{k1} \nnb \\
  &&=-2\epsilon^3 v_1^2 v_2 s_d  (B^u_2)_{12} (B^u_1)_{31} g^u_1 h^u_2 h^u_3,\label{v1square-v2-3}
  \eea
  where we have used $(Y^u_1)_{3k}=(Y^u_2)_{1i}=0$.  Comparing with (\ref{v1square-v2-2}), (\ref{v1square-v2-3}) is at higher order and
  can be neglected. 
  \vskip 0.5cm
  
  Determinant of $M^{'d}$ in (\ref{Mmatrix2}) is calculated as follows.\\
  $\det(M^{'d})$ can be computed using  ${\hat M}^{'d}=K^\dagger_{13} M^{'d}$
 \bea
 {\hat M}^{'d}={\hat X}^d_1 {\hat Y}^d_1 +\epsilon {\hat Y}^d_1 C^d v^*_1 
 +{\hat X}^d_2 {\hat Y}^d_2 v^*_2 +\epsilon {\hat Y}^d_2 C^d v^*_2,
 \eea
 where ${\hat X}^d_{1,2}=K_{13}^\dagger {\hat X}^d_{1,2} K_{13}= 1-\epsilon A^d_{1,2}+\epsilon {\hat B}^d_{1,2}$
 with $ {\hat B}^d_{1,2}=K_{13}^\dagger  ~B^d_{1,2} ~K_{13}$, ${\hat Y}^d_{1,2}=K^\dagger_{13} Y_{12}$.
 As can be seen in (\ref{mixing2}), we would have $({\hat Y}^d_1)_{2i}=({\hat Y}^d_1)_{3i}=({\hat Y}^d_1)_{13}=0$
 and $({\hat Y}^d_2)_{1i}=({\hat Y}^d_2)_{23}=({\hat Y}^d_2)_{31}=({\hat Y}^d_2)_{32}=0$.
 
  4)  $(v^*_1)^3$ term in $\det({\hat M}^{'d})$ comes from $\det(v^*_1 {\hat X}^d_1 {\hat Y}^d_1+\epsilon v^*_1 {\hat Y}^d_1 C^d)$. 
  Two matrices appearing in it are both rank one. So the determinant of this $3\times 3$ matrix must be zero.

 5)  $(v^*_1)^2 v^*_2$ term in $\det({\hat M}^{'d})$ should have contributions from both $v^*_1 {\hat X}^d_1 {\hat Y}^d_1$ 
 and $\epsilon v^*_1 {\hat Y}^d_1 C^d$
 since they are both rank one. Since $({\hat Y}^d_1 C^d)_{2i}=({\hat Y}^d_1 C^d)_{3i}=0$
 and $({\hat Y}^d_1)_{13}=({\hat Y}^d_2)_{23}=({\hat Y}^d_1 C^d)_{13}=0$, 
 this part of the determinant is
 \bea
 &&\epsilon (v^*_1)^2 v^*_2 \varepsilon^{ijk}  ({\hat Y}^d_1 C^d)_{1i}[ ({\hat X}^d_1 {\hat Y}^d_1)_{2j} 
 ({\hat X}^d_2 {\hat Y}^d_2 +\epsilon {\hat Y}^d_2 C^d)_{3k}
 + ({\hat X}^d_2 {\hat Y}^d_2 +\epsilon {\hat Y}^d_2 C^d)_{2j} ({\hat X}^d_1 {\hat Y}^d_1)_{3k} ] \nnb \\
 &&~~=\epsilon^2 (v^*_1)^2 v^*_2 \varepsilon^{ijk}  ({\hat Y}^d_1 C^d)_{1i}
 [ ({\hat B}^d_1)_{21} ({\hat Y}^d_1)_{1j}  ({\hat X}^d_2 {\hat Y}^d_2 +\epsilon {\hat Y}^d_2 C^d)_{3k}
 + ({\hat X}^d_2 {\hat Y}^d_2 +\epsilon {\hat Y}^d_2 C^d)_{2j} ({\hat B}^d_1)_{31} ({\hat Y}^d_1)_{1k}] \nnb \\
 &&~~\approx \epsilon^2 (v^*_1)^2 v^*_2 \varepsilon^{ijk}  ({\hat Y}^d_1 C^d)_{1i}
 [ ({\hat B}^d_1)_{21} ({\hat Y}^d_1)_{1j}  ({\hat Y}^d_2 )_{3k}
 + ( {\hat Y}^d_2 )_{2j} ({\hat B}^d_1)_{31} ({\hat Y}^d_1)_{1k}] \nnb \\
 &&~~=\epsilon^2 (v^*_1)^2 v^*_2 \varepsilon^{ijk} ({\hat B}^d_1)_{21} 
 [ ({\hat Y}^d_1 C^d)_{11} ({\hat Y}^d_1)_{12}- ({\hat Y}^d_1 C^d)_{12} ({\hat Y}^d_1)_{11}] ({\hat Y}^d_2)_{33} \nnb \\
 && ~~=\frac{1}{2}\epsilon^2 (v^*_1)^2 v^*_2 s_{23}c_{23}s_{13}s_{12}c_{12} h^d_3 h^d_2 g^d_1 (g^d_2 h^d_2 -g^d_1 h^d_1)
 [(h^u_3)^2-(h^u_2)^2].
 \label{v1square-v2-4}
 \eea
 In the fourth line in (\ref{v1square-v2-4}) we have used the fact that the factors times $({\hat B}^d_1)_{31}$
 is zero since if any one of the $i,j,k$ indices equals to 3, an associated factor would be zero.

 6) $(v^*_2)^3$ term comes from $\det( v^*_2 {\hat X}^d_2 {\hat Y}^d_2 +v^*_2 \epsilon {\hat Y}^d_2 C^d)$.
 The leading term in it is $\det[(1-\epsilon A^d_2) {\hat Y}^d_2 +\epsilon {\hat Y}^d_2 C^d]=
 \det[{\hat Y}^d_2 (1-\epsilon A^d_2гл\epsilon C^d) ]=\det({\hat Y}^d_2) \det(1-\epsilon A^d_2+\epsilon C^d)=0$.
 Since $({\hat Y}^d_2)_{1i}=0$, $({\hat Y}^d_2)_{31}=({\hat Y}^d_2)_{32}=0$ and 
 $({\hat Y}^d_2 C^d)_{31}=({\hat Y}^d_2 C^d)_{32}=0$, the leading non-zero term is
 \bea
 &&\epsilon^2 (v^*_2)^3 \varepsilon^{ijk}({\hat B}^d_2 {\hat Y}^d_2)_{1i} [
 ({\hat Y}^d_2)_{2j} ({\hat Y}^d_2 C^d)_{3k}+ ({\hat Y}^d_2 C^d)_{2j}  ({\hat Y}^d_2)_{3k}] \nnb \\
 &&~~~=\epsilon^2 (v^*_2)^3 \varepsilon^{ijk}[({\hat B}^d_2)_{13}({\hat Y}^d_2)_{3i} ({\hat Y}^d_2)_{2j} ({\hat Y}^d_2 C^d)_{3k}
 +({\hat B}^d_2)_{12}({\hat Y}^d_2)_{2i} ({\hat Y}^d_2 C^d)_{2j}  ({\hat Y}^d_2)_{3k}] \nnb \\
 &&~~~=\epsilon^2 (v^*_2)^3 \varepsilon^{ijk} [ ({\hat B}^d_2)_{13} ({\hat Y}^d_2 C^d)_{3k}
 +({\hat B}^d_2)_{12} ({\hat Y}^d_2 C^d)_{2k} ] ({\hat Y}^d_2)_{3i} ({\hat Y}^d_2)_{2j} \nnb \\
 &&~~~=\epsilon^2 (v^*_2)^3 \varepsilon^{3jk} ({\hat B}^d_2)_{12} ({\hat Y}^d_2)_{2k'} (C^d)_{k'k} ({\hat Y}^d_2)_{33} ({\hat Y}^d_2)_{2j}\nnb \\
 &&~~~=\frac{1}{2} \epsilon^2  (v^*_2)^3 s_{13} c_{12} s_{12} c_{23} s_{23} g^d_1 h^d_2 h^d_3 [(h^u_3)^2-(h^u_2)^2]
 (g^d_2 h^d_2 -g^d_1 h^d_1),\label{v2-cubic-2}
 \eea
 where in the fourth line we have used $\varepsilon^{ijk} ({\hat Y}^d_2 C^d)_{3k} ({\hat Y}^d_2)_{3i}=0$.

\end{document}